\documentclass{jps-cp}
\usepackage{txfonts} 

\title{$r$-Process and Kilonovae}

\author{Shinya \textsc{Wanajo}$^{1, 2, 3}$}

\inst{$^{1}$Max-Planck-Institut f\"ur Gravitationsphysik (Albert-Einstein-Institut), Am M\"uhlenberg 1, D-14476 Potsdam-Golm, Germany \\
$^{2}$ RIKEN iTHEMS Research Group, 2-1 Hirosawa, Wako, Saitama 351-0198, Japan\\
$^{3}$ Department of Engineering and Applied Sciences, Faculty of Science and Technology, Sophia University, 7-1 Kioicho, Chiyoda-ku, Tokyo 102-8554, Japan}
\email{shinya.wanajo@aei.mpg.de}

\recdate{September 13, 2019}

\abst{The kilonova associated with the neutron star merger GW170817
provides us with several hints to elucidate the nature of the
$r$-process in the universe. In this article, we inspect the radioactive
isotopes that powered the kilonova emission, provided that the merger
ejecta consisted of material with a solar $r$-like abundance pattern. It
is suggested that the early (1--10 days after merger) kilonova emission
is mainly due to the $\beta$-decay chain $^{66}$Ni $\rightarrow$
$^{66}$Cu $\rightarrow$ $^{66}$Zn, which can be the source of the
steepening of the light curve at about 7 days. The late time ($> 10$
days) heaing is attributed to the $\alpha$-decay and fission
($^{254}$Cf) of trans-Pb species (in addition to $\beta$-decay), which
can be the signature of $r$-processing beyond the heaviest stable
elements. This article summarizes a recent work by the author
\cite{Wanajo2018} with the additional calculations of kilonova light
curves by using a numerical code in \cite{Hotokezaka2019}.}

\kword{nucleosynthesis, neutron star mergers, kilonovae, r-process}

\begin{document}
\maketitle

\section{Introduction}
\label{sec:intro}

The discovery of a kilonova \cite{Coulter2017, Valenti2017}, the
electromagnetic counterpart of the gravitational wave signal from the
neutron star merger GW170817 \cite{Abbott2017}, gives us a unique
opportunity to directly inspect the $r$-process nucleosynthesis in the
universe. The luminosity of the kilonova indicates the ejecta mass of
$M_\mathrm{ej}/M_\odot = 0.03$--0.06 \cite{Cowperthwaite2017,
Nicholl2017, Tanaka2017, Kawaguchi2018}, which points an ejection of
material from the post-merger accretion disk \cite{Dessart2009,
Metzger2014, Just2015, Lippuner2017, Shibata2017, Siegel2017,
Fujibayashi2018} in addition to the early dynamical ejecta
\cite{Freiburghaus1999, Goriely2011, Bauswein2013, Wanajo2014,
Sekiguchi2015, Sekiguchi2016, Radice2018}. Its spectral evolution
suggests the lanthanide mass fraction of $X_\mathrm{lan} = 0.001$--0.01
\cite{Arcavi2017, Chornock2017, Nicholl2017, Waxman2018}, which confirms
the production of the heavy elements with $Z \ge 57$ ($A \ge 139$). An
identification of Sr has recently been reported by inspection of the
highly Doppler-shifted spectra of the kilonova ejecta
\cite{Watson2019}. Nevertheless, an evidence of the production of the
heaviest $r$-process elements such as gold and uranium is still
missing. Furthermore, no information of the nucleosynthetic abundance
distribution has been obtained, which in fact determines the radioactive
heating rate for the kilonova emission. Here we try to identify the
dominant radioactive species that powered the kilonova emission of the
merger GW170817. The merger ejecta are assumed to be composed of
material with a solar $r$-like abundance pattern, in light of the
spectroscopic study of $r$-enhanced stars in the Milky Way (MW) halo
\cite{Cowan2019}. This article summarizes the work in \cite{Wanajo2018}
with updates by including the kilonova light curves computed with a
numerical code in \cite{Hotokezaka2019}.

\section{Models and nucleosynthesis}
\label{sec:model}

\begin{table}[tbh]
\caption{Mass fractions (in units of \%) of lanthanides and those of radioactive isotopes at 0.1 days.}
\label{tab:abun}
\begin{tabular}{llllllllllll}
\hline
\hline
Model & $A_\mathrm{min}$ & lanthanide & $^{66}$Ni & $^{72}$Ge & $^{127}$Sn & $^{132}$I & $^{222}$Rn & $^{223}$Ra & $^{224}$Ra & $^{225}$Ra & $^{254}$Cf \\
\hline
mFE-a & 69 & 1.4 & 2.6 & 0.66 & 0.28 & 0.27 & 0.0095 & 0.010 & 0.0057 & 0.017 & 0.0040 \\
mFE-b & 88 & 8.6 & 0.30 & 0.14 & 1.8 & 1.6 & 0.057 & 0.061 & 0.035 & 0.10 & 0.026 \\
\hline
\end{tabular}
\end{table}

\begin{figure}
\vspace{3.5cm}
\hspace{-2cm}
\includegraphics[width=7.8cm]{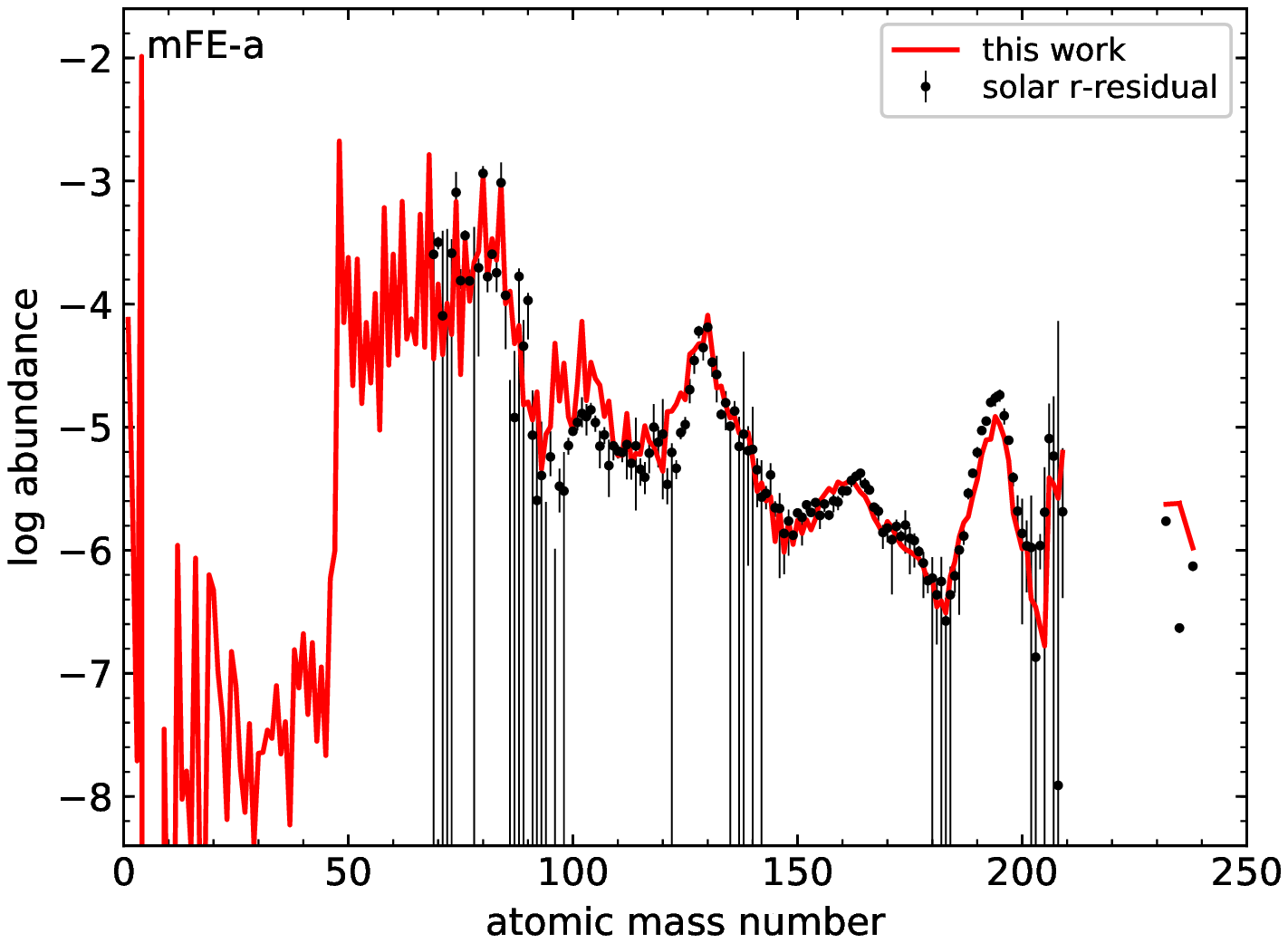}
\includegraphics[width=7.8cm]{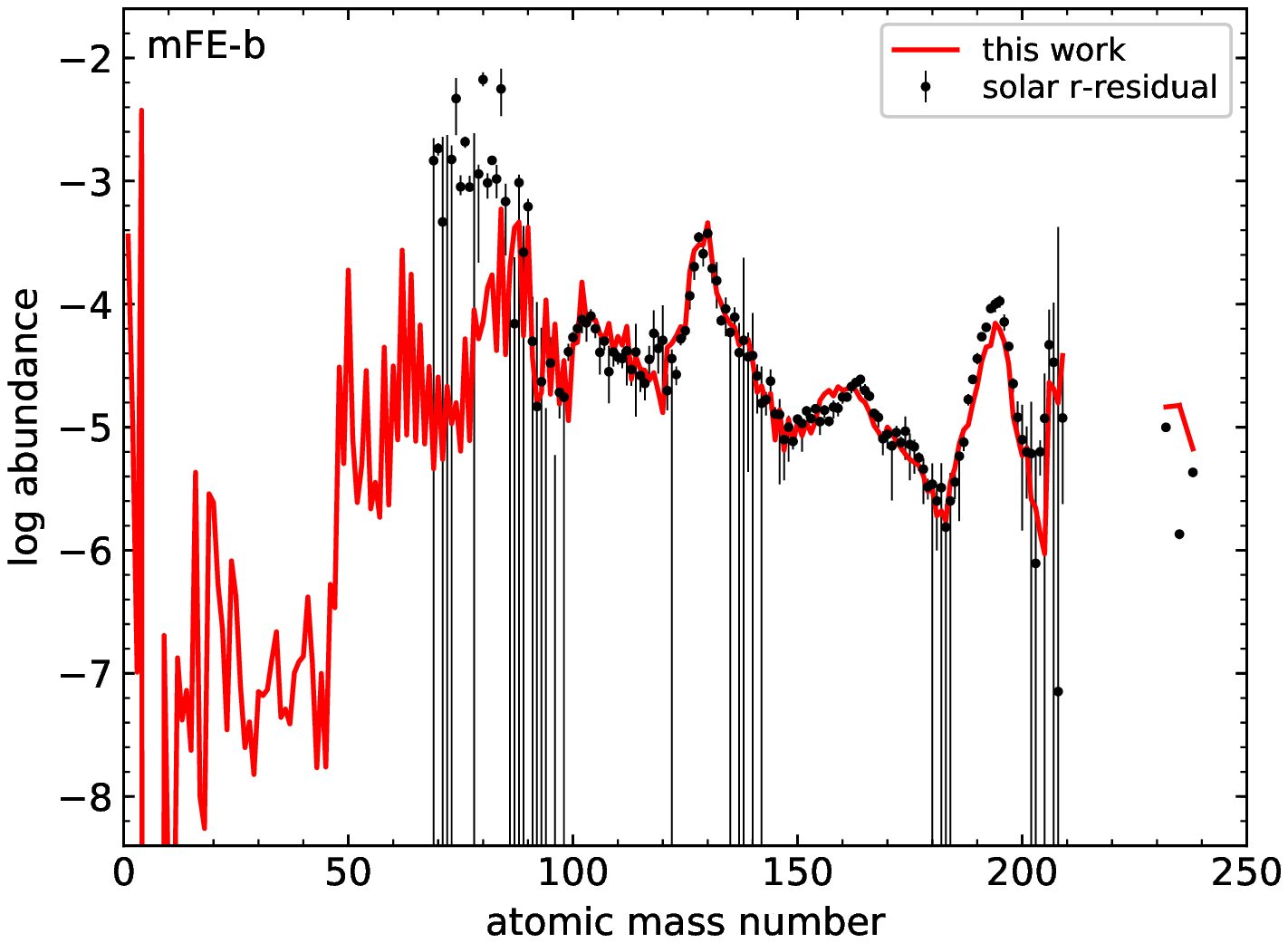}
\vspace{-4.5cm}
\caption{Final nuclear abundances (red lines) for mFE-a (left) and mFE-b (right). The black circles (with error bars) indicate the $r$-residuals to the solar abundances \cite{Goriely1999}, which are vertically shifted to match the abundances at $A = 138$.}
\label{fig:abun}
\end{figure}

Thermodynamic histories of the ejecta are obtained from a free expansion
model (FE) \cite{Wanajo2018}, in which the density and temperature are
calculated from a suite of three parameters, i.e., expansion velocity
($v$), initial entropy ($S$; in units of Boltzmann constant per
nucleon), and initial electron fraction ($Y_\mathrm{e}$). The adopted
ranges of parameters are $(v/c,\, S,\, Y_\mathrm{e}) = (0.05$-0.30,
10-35, 0.01-0.50) with the intervals $(\Delta v/c,\, \Delta S,\, \Delta
Y_\mathrm{e}) = (0.05$, 5, 0.01), 1800 FEs in total, which cover those
predicted from the recent hydrodynamical simulations of neutron star
mergers. Temporal evolution of 6300 isotopes for each FE is computed by
using an up-to-date reaction network code described in
\cite{Wanajo2018}.

In this study, we assume that the ejecta from a neutron star merger
compose of material with a solar $r$-like abundance pattern, according
to the spectroscopic studies of $r$-enhanced stars in the MW halo
\cite{Cowan2019}. The $r$-residuals to the solar system abundances in
\cite{Goriely1999} are taken to be reference for this purpose. Two cases
with different minimum $A$ are considered: a) $A_\mathrm{min} = 69$ and
b) $A_\mathrm{min} = 88$ (Table~\ref{tab:abun}). The reason for these
choices is twofold. First is due to few measurements of elements lighter
than Sr ($A = 88$) in MW halo stars (except for Ga and Ge in several
stars based on only single lines) \cite{Cowan2019, Siqueira2013}. Second
is that the isotopes with $A < 88$ can be made in nuclear statistical
equilibrium (NSE) or nuclear quasi-equilibrium (QSE), a different
condition from an $r$-process \cite{Hartmann1985, Meyer1998,
Wanajo2018b}. For both cases, the maximum $A$ is taken to be
$A_\mathrm{max} = 205$ because of large uncertainties in the
$r$-residuals of Pb and Bi (Fig.~\ref{fig:abun}). Radioactive species Th
and U are not included, either.

For each case, a multi-component FE (mFE) is constructed by fitting a
linear combination of all FEs to the $r$-residuals. The final abundance
distributions (after decay) are in reasonable agreement with those of
the $r$-residuals between $A_\mathrm{min}$ and $A_\mathrm{max}$ for both
cases (mFE-a and mFE-b), as can be seen in
Fig.~\ref{fig:abun}. Moreover, we find that the nuclei with $A <
A_\mathrm{min}$ and $A > A_\mathrm{max}$ are co-produced: Fe-group (for
mFE-a) and trans-Pb nuclei. For mFE-a, the isotopes with $A = 48$-68 are
made with the first-peak abundances of $r$-residuals ($A\sim 80$) under
similar physical conditions (NSE or QSE). The resultant mass fractions
of lanthanides for mFE-a and mFE-b are $X_\mathrm{lan} = 0.014$ and
0.086 (Table~\ref{tab:abun}), respectively. The former (mFE-a) is in
marginal agreement with the inferred value for the kilonova of the
merger GW170817; the latter (mFE-b) is too large to be a reasonable
model for this event.

\section{Radioactive heating rates}
\label{sec:radio}

The heating rates from the decays of radioactive isotopes are computed
for each case (mFE-a or mFE-b) as an ensemble of those in all FEs. The
resulting mass fractions of several important radioactive isotopes at
0.1~days are listed in Table~\ref{tab:abun}.

\subsection{$\beta$-decay}
\label{subsec:beta}

\begin{figure}
\vspace{3.5cm}
\hspace{-2cm}
\includegraphics[width=7.8cm]{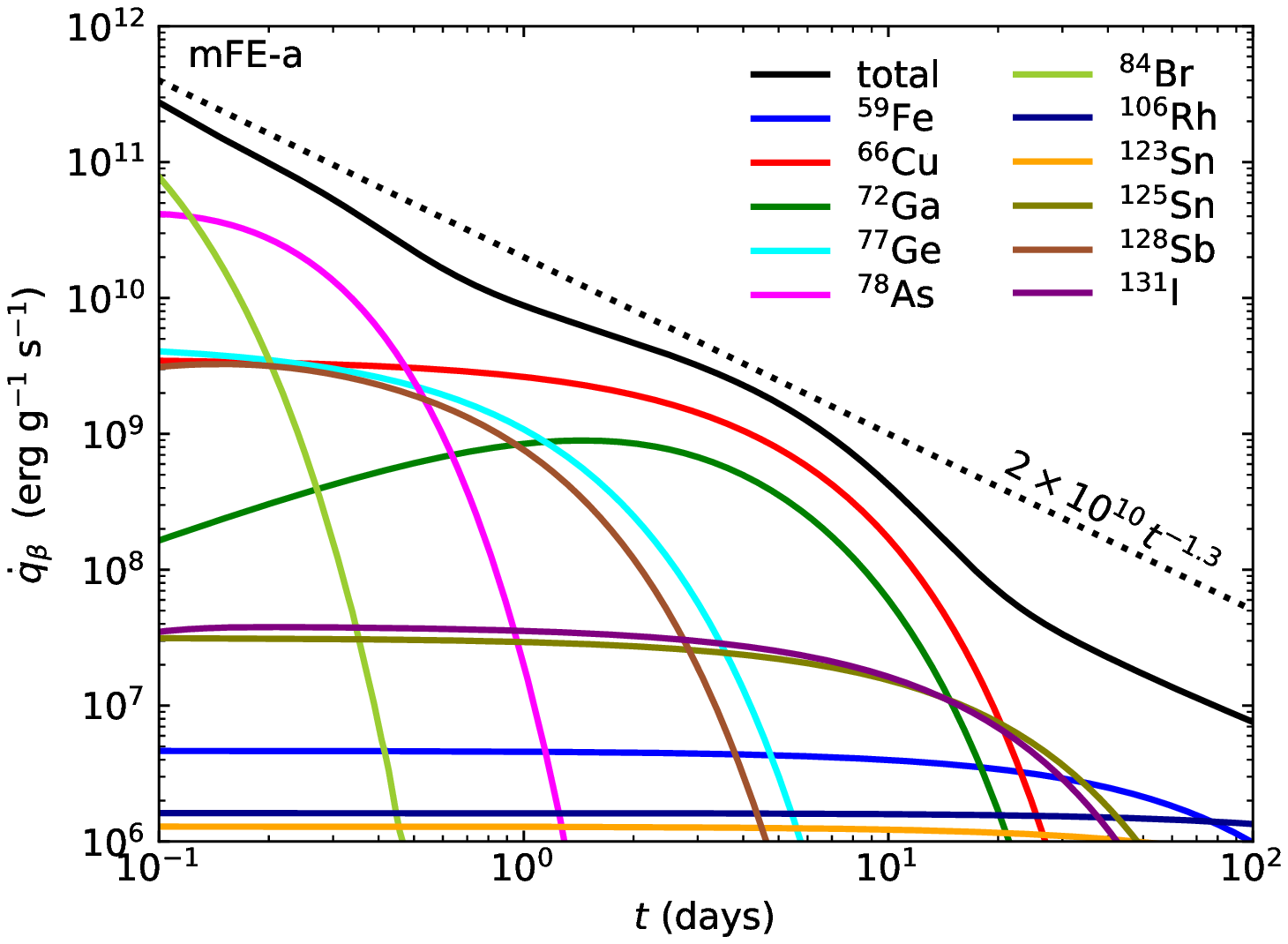}
\includegraphics[width=7.8cm]{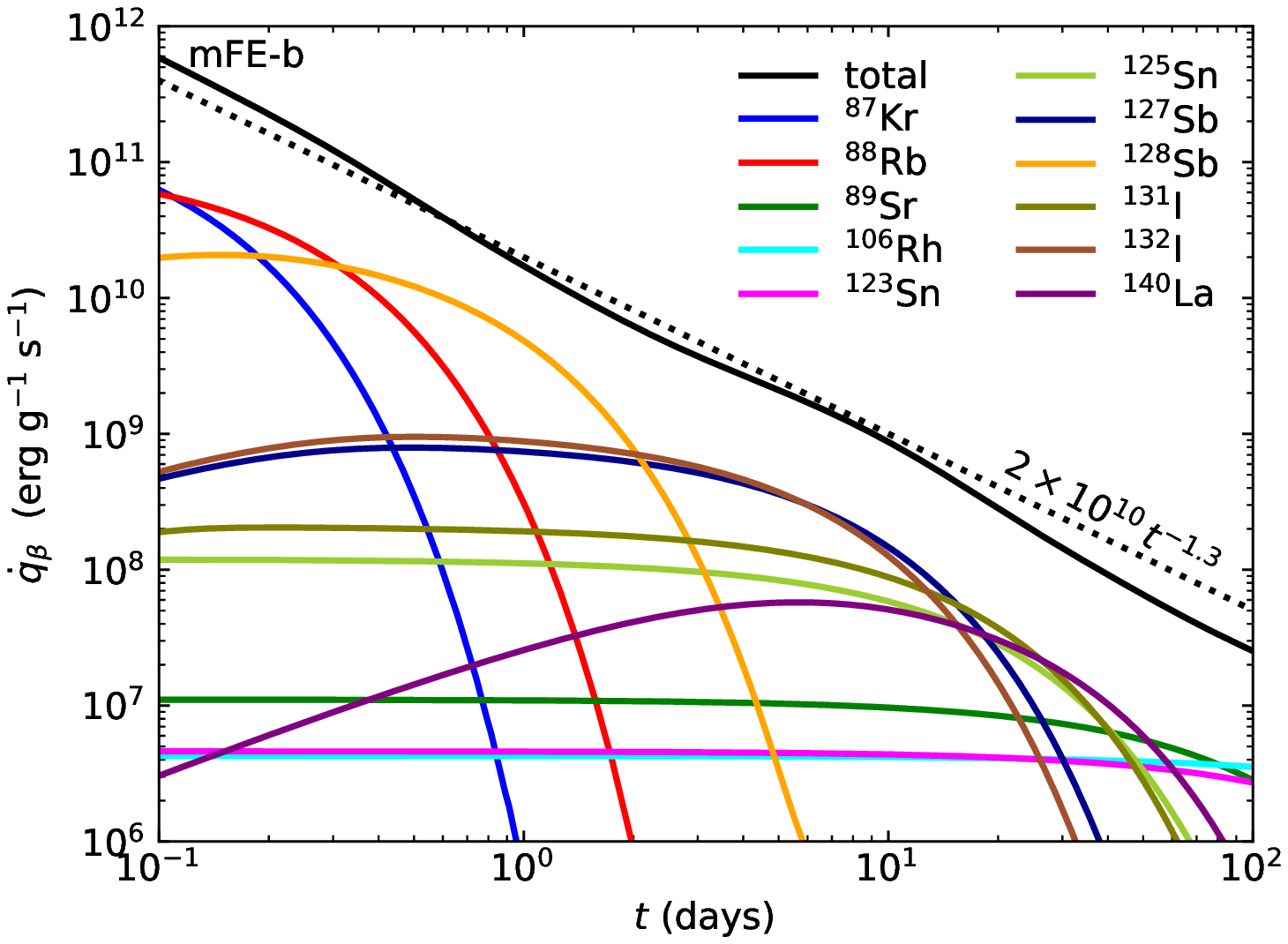}
\vspace{-4.5cm}
\caption{Heating rates from the $\beta$-decays of dominant contributors (solid lines with different colors) for mFE-a (left) and mFE-b (right). The black solid and dotted lines indicate the total heating rate from $\beta$-decay and the empirical (power-law) rate, respectively.}
\label{fig:qbeta}
\end{figure}

The heating rates from $\beta$-decay are shown in Fig.~\ref{fig:qbeta},
where the total and individual rates are indicated by black and colored
lines, respectively. On the one hand, for mFE-b, the total heating rate
from $\beta$-decay approximately follows the power law as suggested in
previous works \cite{Metzger2010, Wanajo2014} because of the
contribution of a number of isotopes around the second peak ($A \sim
130$) with different half-lives. On the other hand, for mFE-a, the total
heating rate exhibits a bump at several days, because of the dominant
decay chain $^{66}$Ni (2.28~d) $\rightarrow$ $^{66}$Cu (5.12~m)
$\rightarrow$ $^{66}$Zn and in part $^{72}$Zn (1.94~d) $\rightarrow$
$^{72}$Ga (14.1~h) $\rightarrow$ $^{72}$Ge; the power law cannot be
taken as an approximation for this case. It should be noted that, for
both mFE-a and mFE-b, these results are robust because of the
experimentally evaluated half-lives and decay energies for the relevant
isotopes (close to the $\beta$-stability).

\subsection{$\alpha$-decay and fission}
\label{subsec:trans-pb}

\begin{figure}
\vspace{3.5cm}
\hspace{-2cm}
\includegraphics[width=7.8cm]{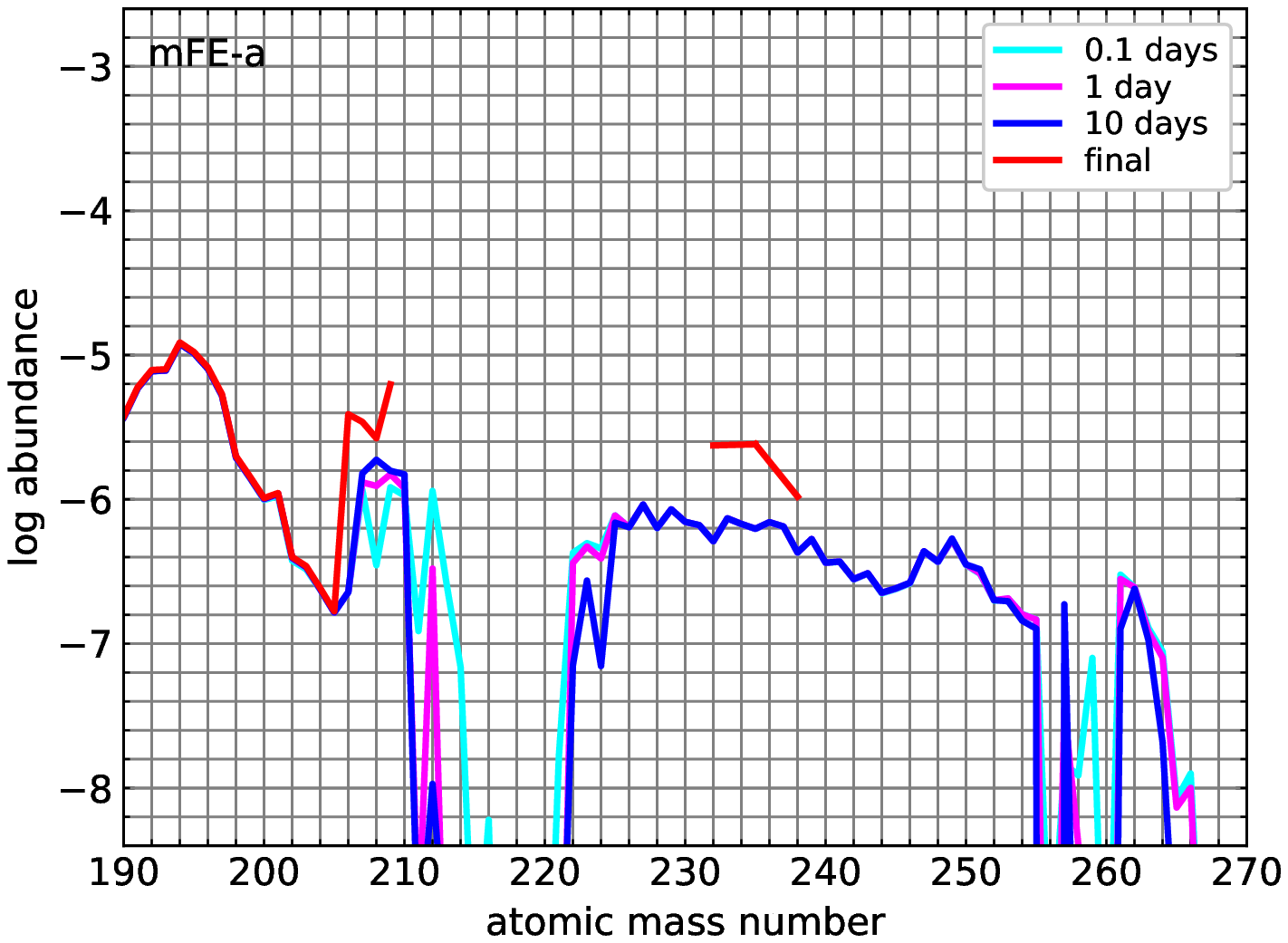}
\includegraphics[width=7.8cm]{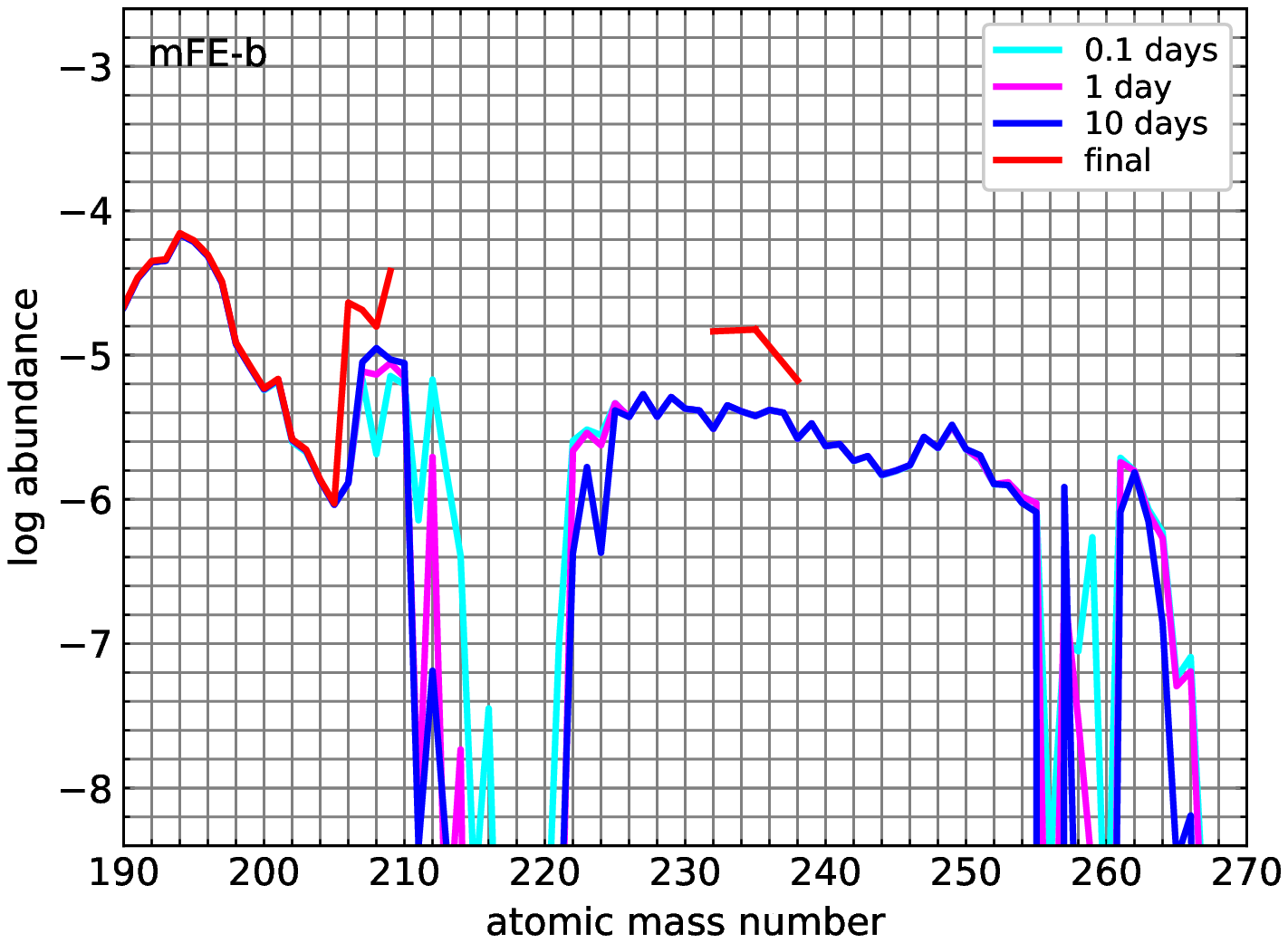}
\vspace{-4.5cm}
\caption{Final nuclear abundances for $A \ge 190$ (red) along with those at 0.1 days (cyan), 1 day (magenta), and 10 days (blue).}
\label{fig:actinide}
\end{figure}

Spontaneous fission and $\alpha$-decay of trans-Pb isotopes are
suggested to be the heating sources of a kilonova at late times ($>
10$~days) \cite{Wanajo2014, Hotokezaka2016, Wanajo2018, Zhu2018,
Wu2019}. Fig.~\ref{fig:actinide} displays the nuclear abundances of
trans-Pb species at different times for mFE-a and mFE-b, both of which
exhibit almost the same abundance patterns but with different amounts
(about 6 times greater for mFE-b). We find that the distribution extends
to $A = 266$ including several Cf and Fm isotopes. It should be noted
that the distribution of trans-Pb isotopes as well as the maximum $A$ is
highly sensitive to the nuclear inputs such as the masses and fission
barriers \cite{Goriely2015, Barnes2016}.

\begin{figure}
\vspace{3.5cm}
\hspace{-2cm}
\includegraphics[width=7.8cm]{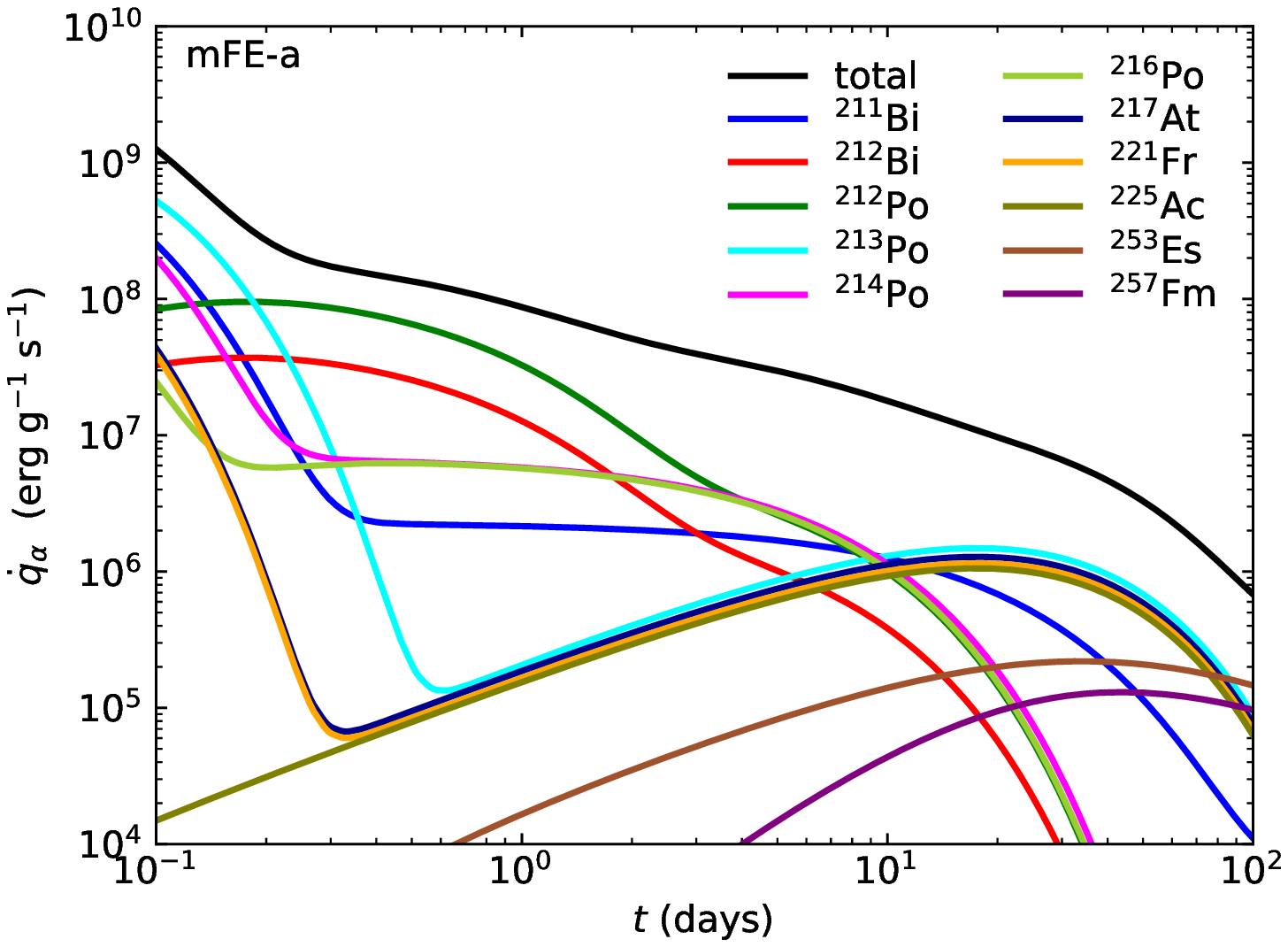}
\includegraphics[width=7.8cm]{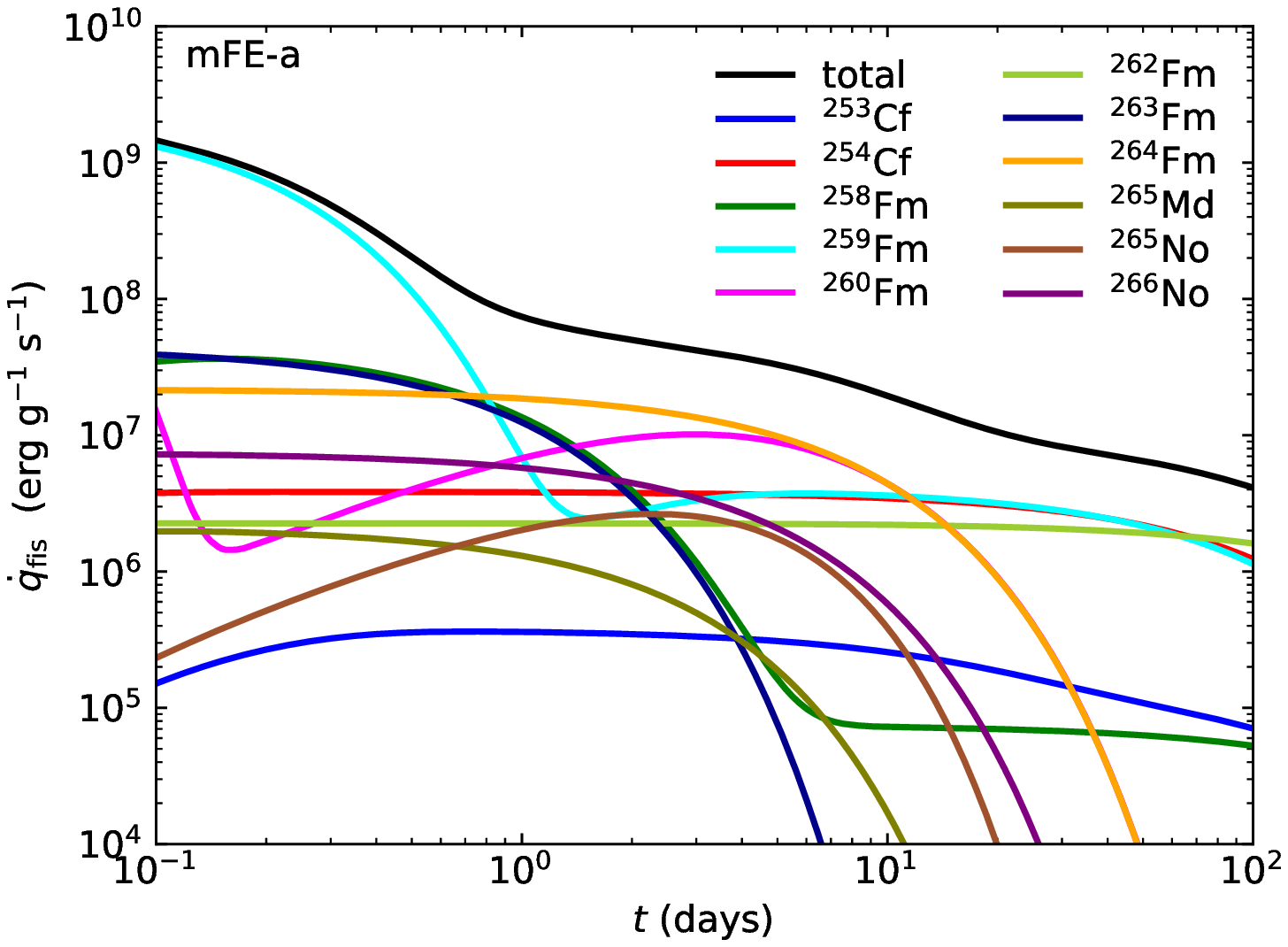}
\vspace{-4.5cm}
\caption{Heating rates from the $\alpha$-decay (left) and fission (right) of dominant contributors (solid lines with different colors) for mFE-a. The black solid line indicates the total heating rate for each decay channel.}
\label{fig:qactinide}
\end{figure}

Fig.~\ref{fig:qactinide} shows the total and individual heating rates
from $\alpha$-decay and fission for mFE-a; those for mFE-b are almost
the same except for their 6 times greater values. The relatively longer
half-lives of dominant contributors than those for $\beta$-decay make
their role important at late times ($> 10$~days) \cite{Wanajo2014,
Hotokezaka2016, Barnes2016}. We find from Fig.~\ref{fig:qactinide}
(left) that the four $\alpha$-decay chains (starting from those in
Table~\ref{tab:abun}) play dominant roles at late times \cite{Wu2019,
Hotokezaka2019}: $^{222}$Rn (3.82~d) $\rightarrow$ $^{218}$Po (3.10~m)
$\rightarrow$ $^{214}$Pb (26.8~m) $\rightarrow$ $^{214}$Bi (19.9~m)
$\rightarrow$ $^{214}$Po (164~$\mu$s) $\rightarrow$ $^{210}$Pb (22.2~yr)
$\rightarrow$ $^{210}$Bi (5.01~d) $\rightarrow$ $^{206}$Tl (4.20~m)
$\rightarrow$ $^{206}$Pb, $^{223}$Ra (11.4~d) $\rightarrow$ $^{219}$Rn
(3.96~s) $\rightarrow$ $^{215}$Po (1.78~ms) $\rightarrow$ $^{211}$Pb
(36.1~m) $\rightarrow$ $^{211}$Bi (2.14~m) $\rightarrow$ $^{207}$Tl
(4.77~m) $\rightarrow$ $^{207}$Pb, $^{224}$Ra (3.66~d) $\rightarrow$
$^{220}$Rn (55.6~s) $\rightarrow$ $^{216}$Po (145~ms) $\rightarrow$
$^{212}$Pb (10.6~h) $\rightarrow$ $^{212}$Bi (1.01~h) $\rightarrow$
$^{208}$Tl (3.05~m) $\rightarrow$ $^{208}$Pb, and $^{225}$Ra (14.9~d)
$\rightarrow$ $^{225}$Ac (10.0~d) $\rightarrow$ $^{221}$Fr (4.77~m)
$\rightarrow$ $^{217}$At (32.3~ms) $\rightarrow$ $^{213}$Bi (45.6~m)
$\rightarrow$ $^{213}$Po (3.72~$\mu$s) $\rightarrow$ $^{209}$Pb (3.25~h)
$\rightarrow$ $^{209}$Bi. For spontaneous fission, $^{254}$Cf (with the
half-live 60.5~days) and a few Fm isotopes become dominant contributors
at late times. The reason for the contribution of Fm isotopes with very
short half-lives (e.g., 1.5~s for $^{259}$Fm) at late times is due to
the theoretically predicted (relatively long) $\beta$-decay half-lives
along their isobars.

\section{Kilonova light curves}
\label{sec:kilonova}

\begin{figure}
\vspace{3.5cm}
\hspace{-2cm}
\includegraphics[width=7.8cm]{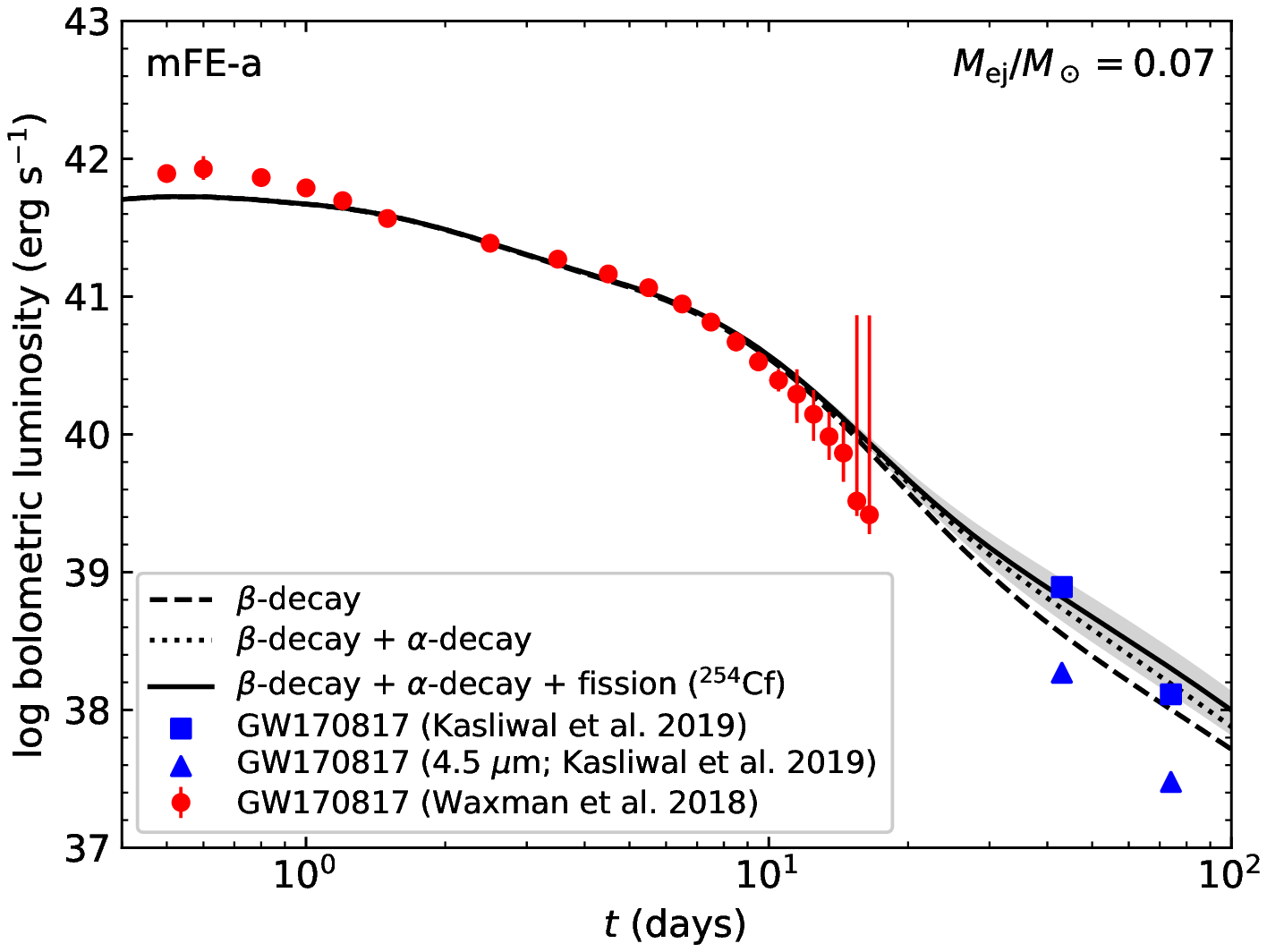}
\includegraphics[width=7.8cm]{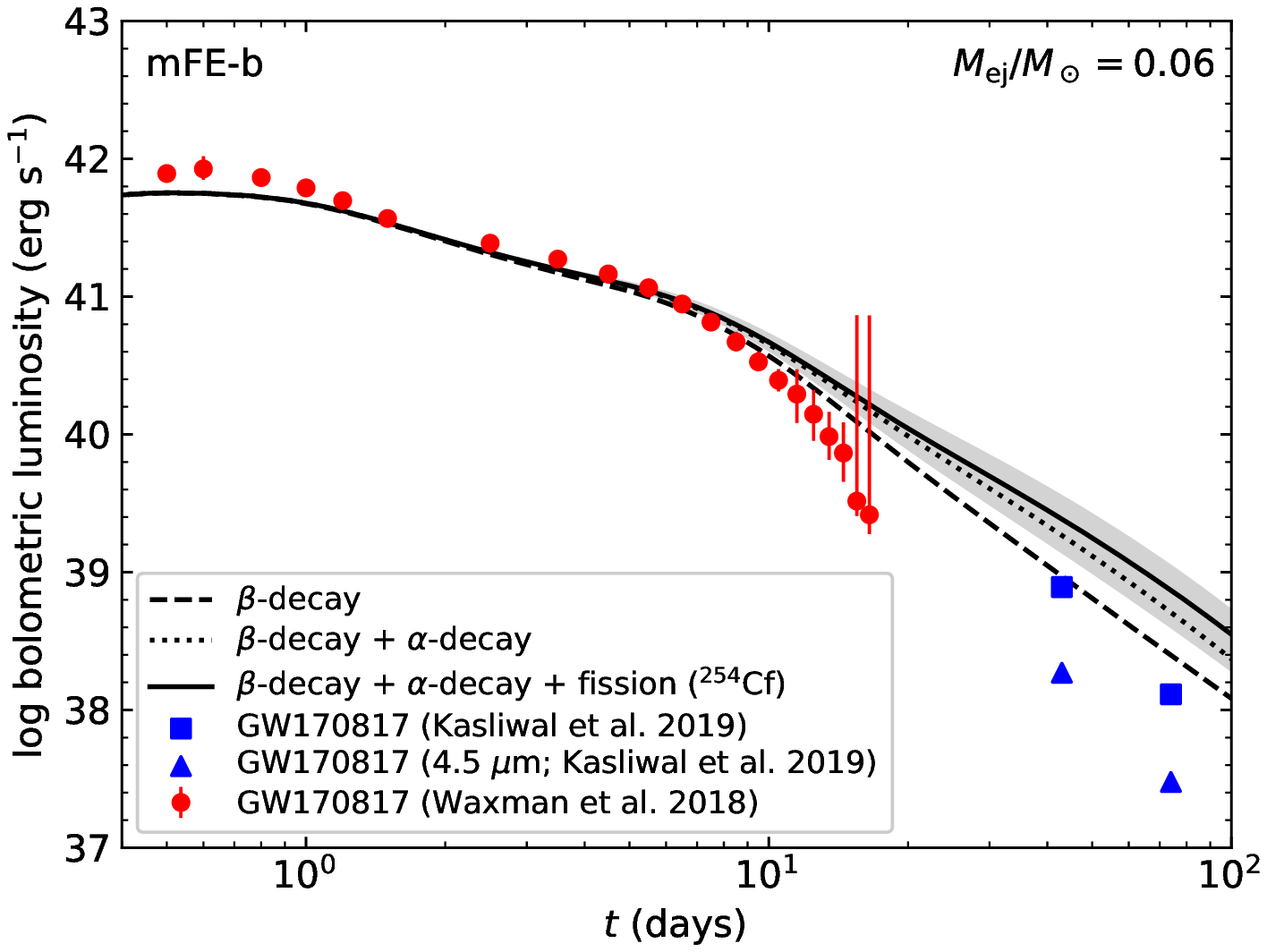}
\vspace{-4.5cm}
\caption{Light curves for mFE-a (left) and mFE-b (right) computed with the numerical code in \cite{Hotokezaka2019}. The assumed ejecta masses are $M_\mathrm{ej}/M_\odot = 0.07$ and 0.06, respectively. For the radial density profile of the ejecta, Eq.~(12) in \cite{Hotokezaka2019}, $v_0/c = 0.1$, $v_\mathrm{max}/c = 0.4$, and $n = 4.5$ are adopted. The opacities (in units of cm$^2$~g$^{-1}$) in mFE-a and mFE-b are chosen to be, respectively, 0.3 and 0.5 for $v > v_\kappa$ and 0.3 (same for both cases) for $v/c < v_\kappa$, where $v_\kappa/c = 0.14$ and 0.20. The solid line indicates the bolometric luminosity (erg~g$^{-1}$) as a function of time (days), in which all the decay channels ($\beta$-decay, $\alpha$-decay, and fission) are included. The gray shaded region brackets the range of the light curve owing to the variation of trans-Pb abundances (see text). The dotted and dashed lines are those without fission ($\beta$-decay and $\alpha$-decay) and without the contribution of trans-Pb nuclei ($\beta$-decay only). Here $^{254}$Cf is taken as the exclusive source of the fission channel. The red circles (with error bars) are the bolometric luminosities of the kilonova associated with GW170817 \cite{Waxman2018}. The blue triangles and squares are, respectively, the luminosities at 4.5 $\mu$m and the inferred bolometric luminosities at 43 days and 74 days after merger \cite{Kasliwal2019}.}
\label{fig:lightcurve}
\end{figure}

The kilonova light curves for mFE-a and mFE-b are computed by using the
numerical code described in \cite{Hotokezaka2019}
(Fig.~\ref{fig:lightcurve}). The ejecta masses are taken to be
$M_\mathrm{ej}/M_\odot = 0.07$ and 0.06, respectively, so that the light
curve matches the inferred bolometric luminosities of the kilonova
between 1~day and 10~days after merger (GW170817) \cite{Waxman2018} (see
the figure caption for other parameters). For $\alpha$-decay, the four
decay chains from $^{222}$Rn, $^{223}$Ra, $^{224}$Ra, and $^{225}$Ra
(Table~\ref{tab:abun}) are included. For spontaneous fission, only
$^{254}$Cf (Table~\ref{tab:abun}) is considered, because the
contribution of other (Fm) isotopes is owing to the highly uncertain
$\beta$-decay half-lives that are theoretically predicted (although such
a possibility cannot be excluded).

We find that, when only $\beta$-decay is included (dashed line), both
mFE-a and mFE-b give consistent results with the kilonova luminosity
between 1~day and 15~days (circles with error bars). The former (mFE-a)
is still in agreement with the observation when including the
contributions of $\alpha$-decay (dotted line) and fission (solid
line). However, mFE-b overpredicts the luminosity at 10-15~days when
$\alpha$-decay and fission are considered. The measurements at 43~days
and 74~days in infrared (4.5~$\mu$m; triangles) \cite{Kasliwal2019},
placing the lower limits to the bolometric luminosities, are consistent
with both cases. The inferred bolometric luminosities (squares) by
\cite{Kasliwal2019}, which should be taken with a caution (estimated
from only a single band; 4.5~$\mu$m), also are in reasonable agreement
with mFE-a (when $\alpha$-decay and fission are included) and mFE-b
(when only $\beta$-decay is considered).

As noted in \S~\ref{subsec:trans-pb}, the abundance distribution of
trans-Pb isotopes and thus their contribution to the heating are highly
sensitive to the inputs of nuclear data. Therefore, we consider a
possible range of the bolometric luminosity at late times as what
follows (see \cite{Wanajo2018} for more detail). We can obtain a
constraint to the range of the ratio Th/Eu from the spectroscopic
studies of MW halo stars (and those in the ultra-faint dwarf
Reticulm~II; 24 stars in total \cite{Holmbeck2018}) among $r$-enhanced
stars. Th and Eu are taken to be representative of trans-Pb and lighter
elements, respectively. The gray shaded region in
Fig.~\ref{fig:lightcurve} represents the obtained range of the
bolometric luminosity when the ratio of trans-Pb/Eu is scaled to
Th/Eu. The overprediction of the light curve at 10-15~days in mFE-b is
still evident when the range of the trans-Pb production is considered,
while that in mFE-a is in agreement with the observation.

\section{Summary}
\label{sec:summary}

We have inspected the radioactive isotopes that dominantly powered the
kilonova associated with the neutron star merger GW170817 by using free
expansion models described in \cite{Wanajo2018}. The ejecta were assumed
to be compose of a) NSE/QSE products ($A = 48$-87) + $r$-process
elements ($A \ge 88$) and b) $r$-process elements only ($A \ge 88$). The
former case (mFE-a) is in good agreement with the inferred mass fraction
of lanthanides as well as the light curve of the kilonova (including its
steepening at about 7~days), while the latter (mFE-b) disagrees with
these observational indications. It is concluded, therefore, the
dominant heating source at early times (1-10~days) was the $\beta$-decay
chain from $^{66}$Ni (and in part, that from $^{72}$Zn) according to the
results for mFE-a. Provided that the Th/Eu ratio was between the range
inferred from $r$-enhanced stars, $\alpha$-decay and spontaneous fission
($^{254}$Cf) substantially contributed to the late time ($> 10$~days)
heating, which could be a signature of the heaviest $r$-element
production in the neutron star merger GW170817.

\bigskip

The author thanks Kenta Hotokezaka for useful discussion and his
numerical code that was used to calculate the kilonova light curves. The
author also thanks the Yukawa Institute for Theoretical Physics at Kyoto
University for fruitful discussions during the workshops YITP-T-18-06
``Nucleosynthesis and electromagnetic counterparts of neutron-star
mergers'' and YITP-T-19-04 ``Multi-Messenger Astrophysics in the
Gravitational Wave Era'', which were useful to complete this work. This
work was supported by the RIKEN iTHEMS Project.


\begin{thebibliography}{9}
\bibitem{Wanajo2018} S. Wanajo, ApJ \textbf{868}, 65 (2018).
\bibitem{Hotokezaka2019} K. Hotokezaka and E. Nakar, arXiv:1909.02581 (2019).
\bibitem{Coulter2017} D. A. Coulter, R. J. Foley, C. D. Kilpatrick, et al., Science \textbf{358}, 1556 (2017).
\bibitem{Valenti2017} S. Valenti, D. J. Sand, S. Yang, et al., ApJL \textbf{848}, L24 (2017).
\bibitem{Abbott2017} B. P. Abbott, R. Abbott, T. D. Abbott, et al., PRL \textbf{119}, 161101 (2017).
\bibitem{Cowperthwaite2017} P. S. Cowperthwaite, E. Berger, V. A. Villar, et al., ApJL \textbf{848}, L17 (2017).
\bibitem{Nicholl2017} M. Nicholl, E. Berger, D. Kasen, et al., ApJL \textbf{848}, L18 (2017).
\bibitem{Tanaka2017} M. Tanaka, Y. Utsumi, P. A. Mazzali, et al. PASJ \textbf{69}, 102 (2017).
\bibitem{Kawaguchi2018} K. Kawaguchi, M. Shibata, and M. Tanaka, ApJL \textbf{865}, L21 (2018).
\bibitem{Dessart2009} L. Dessart, C. Ott, A. Burrows, S. Rosswog, and E. Livne, ApJ \textbf{690}, 1681 (2009).
\bibitem{Metzger2014} B. D. Metzger and R. Fern\'ández, MNRAS \textbf{441}, 3444 (2014).
\bibitem{Just2015} O. Just, A. Bauswein, R. A. Pulpillo, S. Goriely, and H.-T. Janka, MNRAS \textbf{448}, 541 (2014).
\bibitem{Lippuner2017} J. Lippuner, R. Fern\'ández, L. F. Roberts, et al., MNRAS \textbf{472}, 904 (2017).
\bibitem{Shibata2017} M. Shibata, S. Fujibayashi, K. Hotokezaka, et al., PRD \textbf{96}, 123012 (2017).
\bibitem{Siegel2017} D. Siegel and B. D. Metzger, PRL \textbf{119}, 1102 (2017).
\bibitem{Fujibayashi2018} S. Fujibayashi, K. Kiuchi, N. Nishimura, Y. Sekiguchi, and M. Shibata, ApJ \textbf{860}, 64 (2018).
\bibitem{Freiburghaus1999} C. Freiburghaus, S. Rosswog, and F.-K. Thielemann, ApJL \textbf{525}, L121 (1999).
\bibitem{Goriely2011} S. Goriely, A. Bauswein, and H.-T. Janka, ApJL \textbf{738}, L32 (2011).
\bibitem{Bauswein2013} A. Bauswein, S. Goriely, and H.-T. Janka, ApJ \textbf{773}, 78 (2013).
\bibitem{Wanajo2014} S. Wanajo, Y. Sekiguchi, N. Nishimura, et al., ApJL \textbf{789}, L39 (2014).
\bibitem{Sekiguchi2015} Y. Sekiguchi, K. Kiuchi, K. Kyutoku, and M. Shibata, PRD \textbf{91}, 064059 (2015).
\bibitem{Sekiguchi2016} Y. Sekiguchi, K. Kiuchi, K. Kyutoku, M. Shibata, and K. Taniguchi, PRD \textbf{93}, 124046 (2016).
\bibitem{Radice2018} D. Radice, A. Perego, K. Hotokezaka, et. al., ApJ \textbf{869}, 130 (2018).
\bibitem{Arcavi2017} I. Arcavi, G. Hosseinzadeh, D. A. Howell, et al., Nature \textbf{551}, 64 (2017).
\bibitem{Chornock2017} R. Chornock, E. Berger, D. Kasen, et al., ApJL \textbf{848}, L19 (2017).
\bibitem{Waxman2018} E. Waxman, E. O. Ofek, D. Kushnir, and A. Gal-Yam, MNRAS \textbf{481}, 3423 (2018).
\bibitem{Watson2019} D. Watson1, C. J. Hansen, J. Selsing, et al., Nature \textbf{574}, 497 (2019).
\bibitem{Cowan2019} J. J. Cowan, C. Sneden, J. E. Lawler, et al., arXiv:1901.01410 (2019).
\bibitem{Goriely1999} S. Goriely, A\&A \textbf{342}, 881 (1999).
\bibitem{Siqueira2013} C. Siqueira Mello Jr., M. Spite, B. Barbuy, et al., A\&A \textbf{550}, A122 (2013).
\bibitem{Hartmann1985} D. Hartmann, S. E. Woosley, and M. F. El Eid, ApJ \textbf{297}, 837 (1985).
\bibitem{Meyer1998} B. S., Meyer, T. D., Krishnan, and D. D. Clayton, ApJ \textbf{498}, 808 (1998).
\bibitem{Wanajo2018b} S. Wanajo, B. M\"uller, H.-T. Janka, and A. Heger, ApJ \textbf{852}, 40 (2018).
\bibitem{Metzger2010} B. D. Metzger, G. Mart\'ínez-Pinedo, S. Darbha, et al., MNRAS \textbf{406}, 2650 (2010).
\bibitem{Hotokezaka2016} K. Hotokezaka, S. Wanajo, M. Tanaka, et al., MNRAS \textbf{459}, 35 (2010).
\bibitem{Zhu2018} Y. Zhu, R. T. Wollaeger, N. Vassh, et al., ApJL \textbf{863}, L23 (2018).
\bibitem{Wu2019} M.-R. Wu, J. Barnes, G. Mart\'inez-Pinedo, and B. D. Metzger, PRL \textbf{122}, 062701 (2019).
\bibitem{Goriely2015} S. Goriely, EPJA \textbf{51}, 22 (2015).
\bibitem{Barnes2016} J. Barnes, D. Kasen, M.-R. Wu, and G. Mart\'ínez-Pinedo, ApJ \textbf{829}, 110 (2016).
\bibitem{Kasliwal2019} M. M. Kasliwal, D. Kasen, R. M. Lau, et al., MNRAS, in press (2019).
\bibitem{Holmbeck2018} E. M. Holmbeck, T. C. Beers, I. U. Roederer, et al., ApJL \textbf{859}, L24 (2018).
\end{thebibliography}
\end{document}